\documentclass{rspublic} 
\usepackage{graphicx} 
\def\msun{{\rm M_{\odot}}}

\begin{document} 
 
\title
[Gamma--ray Burst Models]
{Gamma--ray Burst Models}

\author[Andrew King]{Andrew King} 
 
\affiliation{Department of Physics and Astronomy, University of Leicester,  
University Road, Leicester, LE1 7RH, UK}

\maketitle

\begin{abstract}{Gamma-ray bursts, black holes, neutron stars, white dwarfs} 

I consider various possibilities for making gamma--ray bursts,
particularly from close binaries. In addition to the much--studied
neutron star + neutron star and black hole + neutron star cases
usually considered good candidates for short--duration bursts, there
are other possibilities. In particular neutron star + massive white
dwarf has several desirable features. These systems are likely to
produce long--duration GRBs, in some cases definitely without an
accompanying supernova, as observed recently. This class of burst
would have a strong correlation with star formation, and occur close
to the host galaxy. However rare members of the class need not be near
star--forming regions, and could have any type of host galaxy. Thus a
long--duration burst far from any star--forming region would also be a
signature of this class. Estimates based on the existence of a known
progenitor suggest that this type of GRB may be quite common,
in agreement with the fact that the absence of a supernova can only be
established in nearby bursts.

\end{abstract}

\section{Introduction} 

Accretion on to a black hole or neutron star is the most efficient way
of extracting energy from normal matter. Since GRBs are briefly the
brightest objects in the Universe, accretion on to one or other of
these objects must be involved in making them.  Constructing GRB
models then becomes a matter of arranging that a suitable mass falls
on to the accretor on the right timescale.

The total emitted energy of GRBs implies that a significant fraction
of a stellar mass is involved. Since we want the burst to be sudden,
we presumably have to start with some equilibrium situation involving
a stellar mass, and destabilize it. Observations again tell us the
infall timescale is, at least initially, extremely short. This
suggests that we can pick out potential stellar reservoirs by
considering their dynamical timescales.

In the collapsar picture a rapidly--rotating massive star is both
accretor and reservoir. During core collapse the central regions try
to form a black hole, while the rest of the core accretes on to it and
makes the burst. The dynamical timescale of the core suggests that
this produces a long--duration burst. 

The other generic situation in which a star is disrupted on a very
short timescale is in a close binary. In the rest of this paper I
shall consider the various possibilities in turn.

\section{Binary models for GRBs}

The reasoning of the last section implies that the accretor (star 1)
in any binary GRB model must be a neutron star or black hole. There
are various possibilities for the reservoir or donor (star 2), which
largely determine the timescale of the GRB. For simplicity I assume
that the binary orbit is circular and the stars corotate with it
(things are similar even if these assumptions fail). Then Roche
geometry shows that mass transfer obeys the relation
\begin{equation}
{\dot M_2\over M_2} = {1\over D}{\dot J\over J}.
\label{mdot}
\end{equation}
(For derivations of the relations in this Section see e.g. the review
by King (1988) 
Here $J$ is the orbital angular momentum, and the main external driven
of angular momentum loss $\dot J < 0$ is by gravitational radiation
(see below). The quantity $D$ is (\ref{mdot}) is given by
\begin{equation}
D = {5\over 6} + {\zeta\over 2} - {M_2\over M_1}
\label{d}
\end{equation}
where $\zeta$ is the mass--radius index of star 2, i.e. this behaves
as $R_2 \propto M_2^{\zeta}$ during mass transfer. If for example star
2 is a low--mass white dwarf, we have $\zeta = -1/3$ and the star
expands as it loses mass. The importance of $D$ is that its sign gives
the sign of $\dot R_L - \dot R_2$, i.e. the relative motion of the
Roche lobe (determined by the orbital evolution as mass is exchanged,
i.e. on a dynamical timescale) and the stellar radius (determined by
its response to mass loss).

Generally $D$ is $O(1)$. If it is positive (called stable mass
transfer) we have
\begin{equation}
{\dot M_2\over M_2} \sim {\dot J\over J},
\label{mdot2}
\end{equation}
i.e. mass transfer on the angular momentum loss timescale $t_J =
|J/\dot J|$, which for gravitational radiation is
\begin{equation}
t_{\rm GR} = {5\over 32}{c^5\over G^3}{a^4\over M_1M_2(M_1+M_2)},
\label{gr}
\end{equation}
where $a$ is the orbital separation. Now since star 2 fills its Roche
lobe its radius is comparable to the orbit, i.e. $R_2\sim a/$few, and
so
\begin{equation}
t_{\rm GR} \sim 10^{10}\biggl({R_2\over R_{\rm MS}}\biggr)^4~{\rm yr}
\label{gr2}
\end{equation}
where the stellar radius $R_2$ is scaled in terms of the
main--sequence value $R_{\rm MS}$. We immediately find that $t_{\rm
  GR} \sim {\rm few} \times 100~{\rm yr}$ for a white dwarf donor, and 
$t_{\rm GR} \sim 10^{-3}$~s for a neutron--star donor.

Thus even stable mass transfer from a neutron--star donor produces
mass flow rates of a suitable order to explain some gamma--ray
bursts. The dynamical timescale (by Roche geometry very close to the
orbital period $P$) for the donor is also milliseconds. We conclude
that NS + NS and BH + NS binaries are candidates for producing
short--duration GRBs.

\section{Dynamical instability}

However stable mass transfer from a neutron--star donor is not the
only possibility for making GRBs from close binaries. In some cases
instead of the stable evolution with $D>0$ seen above we have $D <
0$. This means that the Roche lobe cuts into the star on a dynamical
timescale. This is clearly an unstable feedback process, and the mass
transfer rate tries to increase towards the value
\begin{equation}
\dot M_{\rm dyn} \sim {M_2\over P}.
\label{dyn}
\end{equation}
This is evidently a promising situation for making a GRB. There are
essentially three ways it can come about.

\subsection{Violently expanding donor}

If $\zeta < -5/3$, i.e. the donor expands strong on mass loss, eqn
(\ref{d}) shows we have $D < 0$ for any mass ratio $M_2/M_1$. An
example where this may occur is when a neutron star is reduced by mass
transfer to a mass where it wants to become a white dwarf, typically
about $0.2\msun$. This effect may lead to late flares in a BH + NS
binary GRB (Davies et al., 2005), which would have short duration.

\subsection{Very rapid mass transfer}

The expression (\ref{d}) for $D$ assumes that tides acting within the
binary feed back all the angular momentum of the transferred mass to
the donor star. However it is quite likely that this assumption fails
once mass transfer becomes rapid, simply because the tides cannot
cope. In this situation the expression for $D$ gets an extra term
\begin{equation}
D = {5\over 6} + {\zeta\over 2} - {M_2\over M_1} - \nu
\label{d2}
\end{equation}
(e.g. King \& Kolb, 1995) with $\nu = O(1) > 0$. This obviously allows
the possibility that $D < 0$. This means that even mass transfer from
white dwarf donors can become unstable and produce a GRB, and the mass
transfer timescale $\sim 100$~yr is much shorter than the likely tidal
time. Since the white dwarf has a dynamical time of $\sim 10$~s, this
would be a long--duration GRB.

\subsection{Large mass ratio}

For a white dwarf we have $\zeta = -1/3$ so 
\begin{equation}
D = {2\over 3} - {M_2\over M_1},
\label{d3}
\end{equation}
so mass transfer is dynamically unstable for mass ratios $>
2/3$. Since for a white dwarf we must have $M_2 < 1.4\msun$ (the
Chandrasekhar mass), $D < 0$ requires $M_1 < 2.1\msun$, so the most
likely case is with a neutron--star primary. If $M_1 = 1.4\msun$, as
is often found for neutron stars, instability requires $M_2 >
0.9\msun$. We see that NS + massive WD is a good candidate for
producing long--duration GRBs.

Before turning to these systems, two remarks are in order. First, some
binaries may be unstable in more than one of the ways described
above. For example NS + NS binaries are unstable in all three ways (cf
Rosswog et al., 2003; 2004). Second, it is important to remember that
dynamical instability means that the donor star is no longer a
star. Intuition based on the few--body problem may be a poor guide to
what happens, and a full fluid dynamical calculation is required.

\section{GRBs from NS + massive WD}

In the last subsection I showed that a NS+WD binary with $M_2/M_1 >
2/3$ is a good candidate for producing long--duration GRBs. Here I
mention a few desirable properties of these systems. A full paper will
be submitted for publication shortly.

First, such systems exist. The most promising object is PSR
J1141--6545 (Kaspi et al., 2000) which has a 5~hr orbit and will merge
in about $10^9$~yr under the effect of gravitational
radiation. Because the orbit is relativistic, limits on the masses are
known to great precision (Bailes et al,2003) as $M_{\rm wd} = 0.986
\pm 0.2 \msun$ and $M_{\rm ns} = 1.30 \pm 0.2 \msun$. This firmly
establishes the mass ratio as $M_{\rm wd}/M_{\rm ns} > 0.73$, making
dynamical instability and a GRB inevitable. There is an evolutionary
scheme for them (Davies et al., 2002), which begins with two
main--sequence stars very close in mass. The more massive evolves
first and leaves a massive WD, transferring its envelope to the
companion, which then becomes more massive than the primary was, and
indeed massive enough to make a neutron star. Davies et al. (2002)
show that the Milky Way birthrate is $5 \times 10^{-5} - 5\times
10^{-4}$~yr$^{-1}$, and that more that half of these systems merge
within $10^8$~yr. This means that they will tend to show some
association with star formation, although a tail of sparser systems
will not. Their space velocities are not high in general (Davies et
al., 2002), so many will be found close to their hosts. 

Clearly these systems would produce long--duration bursts. In at least
some cases (if the WD is massive enough to have ONeMg composition)
they are unlikely to produce a supernova. This may be an explanation
for the recent GRB 060614, which was a long--duration burst where any
supernova was at least 100 times fainter than any observed. Although
as a population NS + massive WD bursts follow the star--formation
history of their hosts, a long--duration burst from a galaxy devoid of
recent star formation is possible. Observation of such a burst would
thus be very interesting.

\section{Acknowledgments} I gratefully acknowledge a Royal Society
Wolfson Research Merit Award.

\end{document}